\documentclass[twocolumn,showpacs,pra]{revtex4}
\usepackage{graphicx}
\usepackage{dcolumn}
\usepackage{bm}

\begin{document}

\title{Suppression or enhancement of the Fulde-Ferrell-Larkin-Ovchinnikov
order in a one-dimensional optical lattice with particle
correlated tunnelling}
\author{B. Wang$^{1,2}$ and L.-M. Duan$^{1}$}
\affiliation{ $^{1}$FOCUS center and MCTP, Department of Physics,
University of Michigan, Ann Arbor, MI 48109 \\
$^{2}$ Condensed Matter Theory Center, Department of Physics,
University of Maryland, College Park, MD 20742 }

\begin{abstract}
We study through controlled numerical simulation the ground state
properties of spin-polarized strongly interacting fermi gas in an
anisotropic optical lattice, which is described by an effective
one-dimensional general Hubbard model with particle correlated
hopping rate. We show that the Fulde-Ferrell-Larkin-Ovchinnikov
(FFLO) type of state, while enhanced by a negative correlated
hopping rate, can be completely suppressed by positive particle
correlated hopping, yielding to an unusual magnetic phase even for
particles with on-site attractive interaction We also find several
different phase separation patterns for these atoms in an
inhomogeneous harmonic trap, depending on the correlated hopping
rate.

\end{abstract}

\maketitle

The recent experimental observation of normal-superfluidity
transition in spin-polarized ultra-cold Fermi gas has triggered
tremendous interest in this system
\cite{Zwierlein06,Zwierlein06b,Shin06,Partridge06,Partidge06b}. In
a spin-polarized gas, the competition between the Cooper pairing
and the mismatch of Fermi surfaces of different spin species could
lead to some exotic quantum phases \cite{FF64,LO65,Sarma63}. On
this regard, an example with particular interest is the
Fulde-Ferrell-Larkin-Ovchinnikov (FFLO) state, which spontaneously
breaks the translational and rotational symmetry in space
\cite{FF64,LO65}. The stability region of the FFLO\ state in three
dimensions is unfortunately very narrow \cite{s1,s2}, and so far
it has not been observed in the cold atomic gas. It is much easier
to observe the FFLO-like state in a one-dimensional (1D), or
quasi-1D gas due to the Fermi surface nesting \cite{Hu,Orso,3}.
The FFLO type of\ state has been found in 1D systems with a number
of theoretical studies on the Hubbard model and its continuous
version, including the bosonization approach \cite{Yangkun01},
the Bethe ansatz solution combined with the mean field method \cite{Hu,Orso}%
, and the numerical simulations based on the density matrix renormalization
group (DMRG) \cite{s5,Tezuka08,Rizzi08} or the quantum Monte-Carlo \cite%
{Batrouni}.

To test the theoretical predictions and observe the FFLO-like states, one
needs to have strongly interacting atomic gas confined in an anisotropic
optical lattice, where the hopping is basically along only one spatial
direction. The strongly interaction introduced by the Feshbach resonance is
important for experimentally achieving the low-temperature phase of the
system. However, this strongly interaction causes additional complications
to the theoretical model Hamiltonian. Because of the multi-band population
and the atomic collision over the neighboring sites, the system cannot be
described a conventional single-band Hubbard model any more. Instead, based
on the analysis of the local Hilbert space structure and the symmetry
argument, it is shown in \cite{s8,Duan07} that the effective Hamiltonian for
strongly interacting fermions in an optical lattice is described by a
general Hubbard model with particle correlated hopping rates.

In this paper, we investigate the possibility to observe the
FFLO-like state in the 1D general Hubbard model (GHM) with
population imbalance. Note that in an exactly 1D configuration,
the state can have only a quasi-long-range (QLO) order with
divergence in the corresponding susceptibility. However, this QLO
order can be stabilized to a true-long-range order in a quasi-1D
configuration with small transverse tunneling \cite{HT}. So we can
still use the leading QLO order to character the phase of the
system. We show that the particle correlated hopping in the
general Hubbard model has significant influence on the FFLO phase.
While a negative particle correlated hopping rate enhances the
stability of the FFLO state, a sufficiently large positive
particle correlated hopping rate can completely suppress the FFLO\
phase and stabilize an unusual spin-density wave (SDW) state (with
quasi-long range antiferromagnetic order) even for the atoms with
on-site attractive interaction. The solution here is based on
controlled numerical simulation with the tensor network algorithm
in the thermodynamical limit \cite{Vidal} (a variation of the DMRG
method \cite{DMRG}). We also give a analysis of the phase
separation pattern for this system in a weak global harmonic trap.
While the mean field approach finds a large region of the FFLO\
state surrounded by a fully paired superfluid state in the wing
\cite{Hu}, the exact numerical simulation shows that the parameter
window for such a region is very narrow. For the case of the
general Hubbard model, more intriguing phase separation patterns
could occur. For instance, with positive particle correlation
hopping rate, one can have a SDW state at the core and a fully
paired superfluid state in the wing, while in between the state of
the system shows some remanence of the FFLO-type order.

For strongly interacting fermions near a wide Feshbach resonance, the
effective Hamiltonian is given by the following general Hubbard model \cite%
{s8,Duan07}
\begin{eqnarray}
H &=&\sum_{i}\left[ Un_{i\uparrow }n_{i\downarrow }-\mu _{\uparrow
}n_{i\uparrow }-\mu _{\downarrow }n_{i\downarrow }\right] \\
&&-\sum_{\left\langle i,j\right\rangle ,\sigma }\left[ t+\delta g\left( n_{i%
\overline{\sigma }}+n_{j\overline{\sigma }}\right) +\delta tn_{i\overline{%
\sigma }}n_{j\overline{\sigma }}\right] a_{i\sigma }^{\dagger }a_{j\sigma
}+H.c.  \nonumber
\end{eqnarray}%
where $n_{i\sigma }\equiv $ $a_{i\sigma }^{\dagger }a_{i\sigma }$ and $%
n_{i}\equiv n_{i\uparrow }+n_{i\downarrow }$ are particle number operators, $%
\mu _{\sigma }$ stands for the chemical potential for the spin-$\sigma $
species, $\left\langle i,j\right\rangle $ denotes the neighboring sites, and
$a_{i\sigma }^{\dagger }$ is the creation operator to generate a fermion on
the site $i$ with the spin index $\sigma $. The symbol $\overline{\sigma }$
stands for $\left( \downarrow ,\uparrow \right) $ given $\sigma =\left(
\uparrow ,\downarrow \right) $. The $\delta g$ and $\delta t$ terms in the
Hamiltonian describe the particle correlated hopping, which come from the
contributions of the multi-band processes and the direct neighboring
coupling of the atoms. As one moves far away from the Feshbach resonance,
the correlated hopping terms vanish and the Hamiltonian in Eq. (1) reduces
to the conventional Hubbard model. However, near the resonance, $\delta g$
and $\delta t$ can be significant compared with the atomic tunneling rate $t$%
. It is interesting to note that a similar form of the Hamiltonian
has also been proposed for the high-Tc cuprate materials
\cite{htGHM}, so the interest in this model is not limited to the
system of the strongly interacting atomic gas. In Eq. (1), we keep
$\mu _{\uparrow }$ and $\mu _{\downarrow }$ different, which
accounts for the population imbalance in the two spin components.

In this work, we consider an anisotropic optical lattice where the hopping
along the $x,y$ directions are suppressed by the potential barriers. For
this effective one-dimensional system, the ground state can be found through
the time-evolving block decimation (TEBD) algorithm \cite{Vidal}. For this
algorithm, first we transfer the fermions to effective spins through the
Jodan-Wigner transformation \cite{WD}. Each site has four possible states,
denoted by $\left\vert i_{s}\right\rangle $ for the $s$th site. The
coefficient $c_{i_{1}\ldots i_{n}}$ of the ground-state wave function $|\Psi
\rangle =\sum_{i_{1}=1}^{d}\cdots \sum_{i_{n}=1}^{d}c_{i_{1}\ldots
i_{n}}|i_{1}\cdots i_{n}\rangle $ is expressed in the matrix product form:
\begin{equation}
c_{i_{1}\ldots i_{n}}=\sum_{\alpha _{1},\ldots \alpha _{n}=1}^{\chi }\Gamma
_{\alpha _{n}\alpha _{1}}^{[1]i_{1}}\Gamma _{\alpha _{1}\alpha
_{2}}^{[2]i_{2}}\Gamma _{\alpha _{2}\alpha _{3}}^{[3]i_{2}}\cdots \Gamma
_{\alpha _{n-1}\alpha _{n}}^{[n]i_{n}},
\end{equation}%
where $\Gamma ^{\lbrack s]i_{s}}$ denotes the matrix associated with site-$s$
with the matrix dimension $\chi $. The lattice is bipartite, and for
calculation in the thermodynamical limit, we assume a translational symmetry
of the matrix for each sublattice \cite{Vidal, WD}. The matrix is optimized
through minimization of the energy with the imaginary time evolution by the
Hamiltonian $H$. To identify different orders, we calculate the real space
spin ($S_{r}$), density ($D_{r}$), and pair ($P_{r}$) correlations and their
Fourier transformations $X_{k}=1/\sqrt{M}\sum_{r=0}^{M-1}X_{r}\cos (kr)$,
where $M$ is the number of sites involved in the transformation and $X$
stands for $S$, $D,$ or $P$ correlation \cite{WD}. The real space
correlation functions are defined by
\begin{eqnarray}
S_{r}^{m} &\equiv &\langle \mathbf{s}_{i}^{m}\mathbf{s}_{i+r}^{m}\rangle
-\langle \mathbf{s}_{i}^{m}\rangle \langle \mathbf{s}_{i+r}^{m}\rangle ,
\nonumber \\
D_{r} &\equiv &\langle n_{i}n_{i+r}\rangle -\langle n_{i}\rangle \langle
n_{i+r}\rangle , \\
P_{r} &\equiv &\langle a_{i\uparrow }a_{i\downarrow }a_{i+r\downarrow
}^{\dagger }a_{i+r\uparrow }^{\dagger }\rangle -\langle a_{i\uparrow
}a_{i\downarrow }\rangle \langle a_{i+r\downarrow }^{\dagger }a_{i+r\uparrow
}^{\dagger }\rangle ,  \nonumber
\end{eqnarray}%
where the spin operators associated with site $i$ is given by $\mathbf{s}%
_{i}^{m}\equiv a_{i\alpha }^{\dagger }\mathbf{\sigma }_{\alpha \beta
}^{m}a_{i\beta }/2$ with $\alpha $ and $\beta =\downarrow $, $\uparrow $ and
$\mathbf{\sigma }^{m}$ ($m=x$,$y$,$z$) standing for the Pauli matrices. In
the calculation, we typically take $\chi =80$ for the matrix dimension and $%
M=100$ for the Fourier transform (the results have been well converged with
the above choice of the parameters).

\begin{figure}[tbp]
\includegraphics [width=8.5 cm] {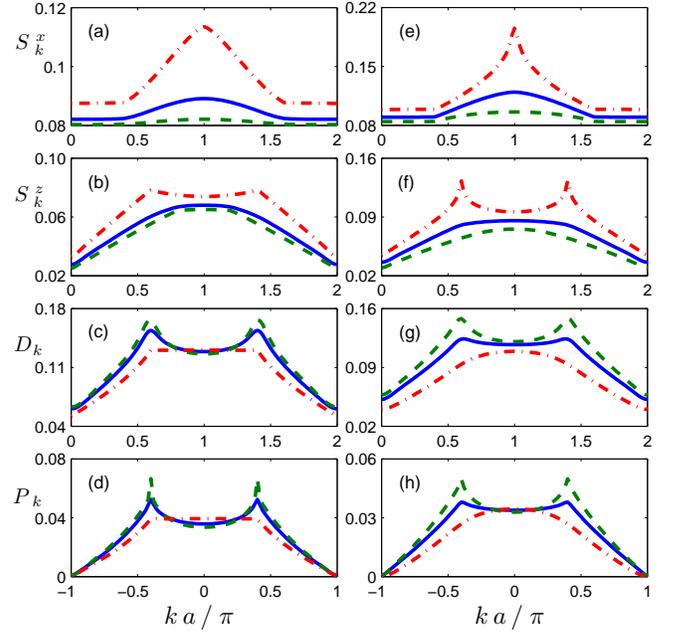}
\caption[{[Fig.}]{Fourier transform of the correlation functions
for the general Hubbard model at half-filling. For plots (a)-(d),
$U=-8t$, while $U=-2t$ for
(e)-(h). For all the plots, the spin polarization is chosen to be the same at $%
p=0.4$. The solid, dashed, and dash-dotted curves correspond to $\protect%
\delta g=0$, $-0.5t$, and $t$, respectively. The parameter $\delta
t$ is taken as $\delta t=-2\delta g$.}
\end{figure}

In Fig. 1, we investigate how the particle correlated hopping rate
in the GHM influences the stability of the FFLO state and the
phase of the system. In our calculation, we take attractive
interaction with $U=-8t$ and $-2t$ respectively for the figures on
the left and the right panels, and choose a fixed population
imbalance with $p=\left( N_{\uparrow }-N_{\downarrow }\right)
/(N_{\uparrow }+N_{\downarrow })=0.4.$ First, for the conventional
Hubbard model with $\delta g=\delta t=0$, the pair correlation
$P_{k}$ peaks at nonzero momenta $k$ with population imbalance,
which is a signature of the FFLO\ like state. The result thus
confirms the previous theoretical predictions. Then, we tune the
particle correlated hopping rate $\delta g$. With a negative
$\delta g$, as shown in Fig. 1, the peaks of $P_{k}$ at non-zero
momenta become sharper, which means that the FFLO order gets
enhanced (the FFLO is the leading quasi-long-range order in this
1D configuration). However, with a positive $\delta g$, the FFLO
peaks in the pair correlation $P_{k}$ get suppressed, and at
$\delta g=t$, the peaks in the spin correlation $S_{k}$ get more
prominent. The peaks in $S_{k}$ are particularly sharp with a
small on-site attraction $U=-2t$, and the spin density wave
emerges as the leading quasi-long range order for the system.
Although the spin density wave order is expected for the repulsive
Hubbard model, its appearance in the case of attractive
interaction is unusual since attraction normally favors some
pairing order. The result of this simulation shows that the
particle correlated hopping can cause some qualitative change to
the phase of the system.

\begin{figure}[tbp]
\includegraphics [width=6 cm] {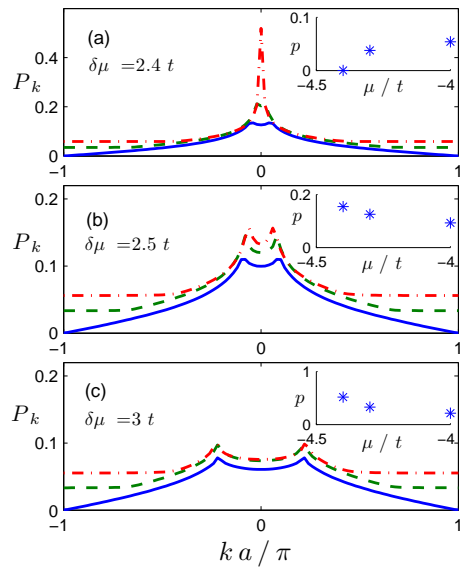}
\caption[{[Fig.}]{Pair correlation functions for the conventional
Hubbard model at $U=-8t$. The difference in chemical potential for
spin-up and
spin-down species are chosen to be $\protect\delta\protect\mu\equiv(\protect%
\mu_{\downarrow}-{\protect\mu_{\uparrow}})/2=2.4t$, $2.5t$, and
$3t$ for plots (a), (b), and (c), respectively. The solid, dashed,
and dash-dotted
lines in the plots correspond to $\protect\mu\equiv (\protect\mu%
_{\uparrow}+\protect\mu_{\downarrow})/2=-4t$ (half-filling), $-4.3t$, and $%
-4.4t$, respectively. The insets show the spin polarization $p$
corresponding to each curve.}
\end{figure}

In the next step, we analyze the possible phase separation
patterns for the polarized fermi gas in an inhomogeneous global
trap $V\left( r\right) $. When the trap $V\left( r\right) $ is
slowly varying from site to site, we can take the local density
approximation where $V\left( r\right) $ basically decreases the
local chemical potential from $\mu $ ($\mu =(\mu _{\uparrow }+\mu
_{\downarrow })/2$) to $\mu -V\left( r\right) $ as one moves from
the trap center to the edge. So to investigate the qualitative
phase separation pattern, we can fix the population imbalance $p$
(and thus also $\delta \mu =(\mu _{\uparrow }-\mu _{\downarrow
})/2$) at the center, and calculate what kind of phases can emerge
as one moves to the trap edge. For the conventional Hubbard model,
the mean-field theory has predicted a large parameter region where
one has a FFLO-like phase at the center, surrounded by a
non-polarizing BCS phase at the edge \cite{Hu}. This separation
pattern is somewhat unusual as in the $3$-dimensional case, one
can only has a BCS phase at the trap center surrounded by other
phases (such as FFLO or polarized normal state) at the edge
\cite{s1,s2}. The recent DMRG calculation however does not find
any evidence of this type of phase separation pattern predicted by
the mean-field theory \cite{s5}. To resolve this problem, we have
performed more extensive calculation over a large parameter
window. We indeed find that this type of phase separation is
possible, however, it exists only for a very narrow parameter
window with the region much smaller than the one predicted by the
mean-field theory.

In Fig. 2, we show different kinds of phase separation patterns for the
conventional Hubbard model with $U=-8t$. We take the density about half
filling at the trap center and vary the population imbalance there. With a
tiny population imbalance at the center (for instance with $p=0.054$ in Fig.
2a, with the corresponding $\delta \mu =2.4t$), when one moves to the edge,
the population imbalance $p$ decreases with decrease of $\mu $ (see the
insert of Fig. 2a), and a non-polarizing BCS state emerges at the edge.
However, when one slightly increase the population imbalance at the center
(for instance with $p=0.093$ in Fig. 2b, with the corresponding $\delta \mu
=2.5t$), the population imbalance increases with a decreasing $\mu $ as one
moves to the edge, and the state finally goes to a polarized normal phase
(with no peaks in the correlation function, not shown in Fig. 2). This is
different from the mean-field prediction \cite{Hu}, where one could get a
non-polarizing BCS state at the edge as long as the central polarization $p$
is below $0.2$. In Fig. 2c, we show the result when the central polarization
$p\approx 0.2$,. Clearly, the polarization is increasing as one moves to the
edge (similar to Fig. 2b), and there is no possibility of a BCS state there.

\begin{figure}[tbp]
\includegraphics [width=8.5 cm] {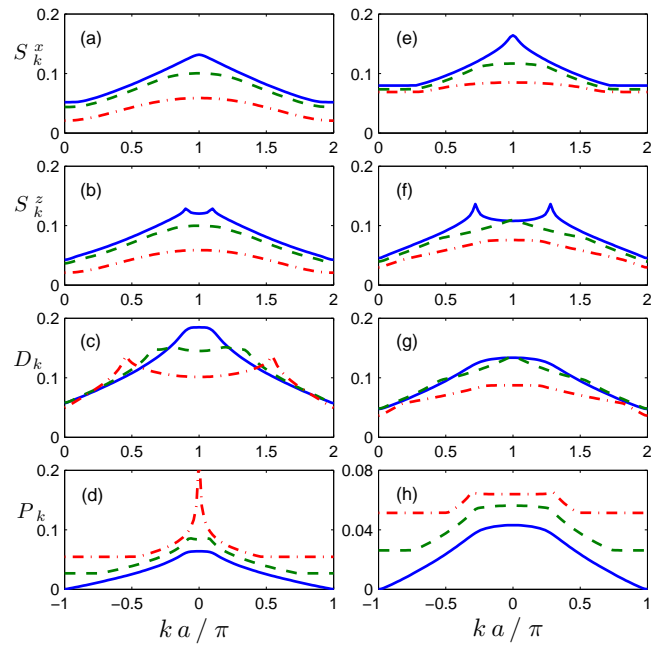}
\caption[{[Fig.}]{The correlation functions for
the general Hubbard model at $U/t=-2$, $\protect\delta g/t=1$ and $\protect%
\delta t/t=-2$. Difference in chemical potentials for spin-up and
spin-down species are chosen to be $\protect\delta\protect\mu\equiv(\protect\mu%
_{\downarrow}-{\protect\mu_{\uparrow}})/2=1.2t$ and $2.4t$ for the
left and right column, respectively. The solid, dashed, and
dash-dotted lines
correspond to $\protect\mu\equiv (\protect\mu_{\uparrow}+\protect\mu%
_{\downarrow})/2=-t$ (half-filling), $-2.5t$, and $-3.3t$,
respectively.}
\end{figure}

In Fig. 3, we investigate the phase separation pattern for the GHM
with particle correlated hopping. The phase separation pattern is
of particular interest for the case of a positive particle
correlated hopping rate $\delta g$. In this case, at the trap
center we have a spin density wave state, while at the edge the
state depends on the overall population imbalance of the system.
In the case of a large population imbalance (the right panel of
Fig. 3), we get a polarized normal state at the edge. However, in
the case of a small population imbalance, the spin-density wave
state can be accompanied by a non-polarizing BCS state at the
edge, as evidenced by the sharp peak in the pair correlation
$P_{k}$ in Fig. 3d. With a negative particle correlated hopping
rate $\delta g$, the phase separation pattern is qualitatively
similar to the case of the conventional Hubbard model, and thus
not shown in Fig. 3.

In summary, through well controlled numerical simulation we have
investigated the properties of the attractive general Hubbard model under
population imbalance, which describes the strongly interacting fermi gas in
a one-dimensional optical lattice. The FFLO type of order gets either
enhanced or suppressed depending on the sign of the particle correlated
hopping rate $\delta g$. When the FFLO\ state is fully suppressed with a
positive $\delta g$, we get an unusual spin density wave state. We also
investigate the phase separation pattern of the system under a weak
inhomogeneous trap, and find several different phase separation patterns
depending on the polarization of the system and the particle correlated
hopping rates.

This work was supported by the MURI, the DARPA, and the DTO under
ARO contracts.


\begin{thebibliography}{99}
\bibitem{Zwierlein06} M. W. Zwierlein, A. Schirotzek, C. H. Schunck, and W.
Ketterle, \textit{Science} \textbf{311}, 492 (2006).

\bibitem{Partridge06} G. B. Partridge et al., \textquotedblleft Pairing and
phase separation in a polarized Fermi gas \textquotedblright , \textit{%
Science} \textbf{311}, 503 (2006).

\bibitem{Zwierlein06b} M. W. Zwierlein, C. H. Schunck, A. Schirotzek, and W.
Ketterle, \textit{Nature} \textbf{442}, 54 (2006).

\bibitem{Shin06} Y. Shin et al., \textit{Phys. Rev. Lett.} \textbf{97},
030401 (2006).

\bibitem{Partidge06b} G. B. Partridge et al., \textit{Phys. Rev. Lett.}
\textbf{97}, 190407 (2006).

\bibitem{FF64} P. Fulde, and R. A. Ferrell, \textit{Phys. Rev.} \textbf{135}%
, A550 (1964).

\bibitem{LO65} A. I. Larkin, and Yu. N. Ovchinnikov, \textit{Sov. Phys. JETP}
\textbf{20}, 762 (1965).

\bibitem{Sarma63} G. Sarma, \textit{J. Phys. Chem. Solids}, \textbf{24},
1029 (1963).

\bibitem{s1} D. E. Sheehy and L. Radzihovsky, Phys. Rev. Lett. 96, 060401
(2006).

\bibitem{s2} W. Zhang, L.-M. Duan, Phys. Rev. A 76, 042710 (2007).

\bibitem{Hu} H. Hu, X.-J. Liu, and P. D. Drummond, \textit{Phys. Rev. Lett.}
\textbf{98}, 070403 (2007).

\bibitem{Orso} G. Orso, \textit{Phys. Rev. Lett.} \textbf{98}, 070402
(2007).

\bibitem{3} M. M. Parish, S. K. Baur, E. J. Mueller, D. A. Huse, Phys. Rev.
Lett. 99, 250403 (2007).

\bibitem{Yangkun01} K. Yang, \textit{Phys. Rev. B} \textbf{63}, 140511(R)
(2001).

\bibitem{s5} A. E. Feiguin and F. Heidrich-Meisner, \textit{Phys. Rev. B}
\textbf{76}, 220508(R) (2007).

\bibitem{Tezuka08} M. Tezuka and M. Ueda, \textit{Phys. Rev. Lett.} \textbf{%
100}, 110403 (2008).

\bibitem{Rizzi08} M. Rizzi et al., \textit{Phys. Rev. B} \textbf{77}, 245105
(2008).

\bibitem{Batrouni} G. G. Batrouni, M. H. Huntley, V. G. Rousseau, and R. T.
Scalettar, \textit{Phys. Rev. Lett.} \textbf{100}, 116405 (2008).

\bibitem{s8} L.-M. Duan, \textit{Phys. Rev. Lett.} \textbf{95}, 243202
(2005).

\bibitem{Duan07} L.-M. Duan, Europhys. Lett. 81, 20001 (2008).

\bibitem{HT} E. W. Carlson, V. J. Emery, S. A. Kivelson, and D. Orgad,
\textquotedblleft Concepts in high temperature
superconductivity\textquotedblright , in \textit{The physics of
superconductors} (Springer, Berlin, 2003).

\bibitem{Vidal} G. Vidal, \textit{Phys. Rev. Lett.} \textbf{98}, 070201
(2007).

\bibitem{DMRG} For a review, see U. Schollwoeck, Rev. Mod. Phys. 77, 259
(2005).


\bibitem{htGHM} M. E. Simon and A. A. Aligia, \textit{Phys. Rev. B} \textbf{48}, 7471
(1993).


\bibitem{WD} B. Wang and L.-M. Duan, \textit{N. J. Phys.} \textbf{10},
073007 (2008).

\end{thebibliography}
\end{document}